\documentclass[aps,pre,twocolumn,groupedaddress,showpacs]{revtex4}
\usepackage{graphicx}
\usepackage{dcolumn}
\usepackage{bm}
\usepackage{epsfig}
\usepackage{color}
\input epsf
\epsfclipon

\begin{document}
\title{Entropy-growth-based model of emotionally charged online dialogues} 
\author{Julian Sienkiewicz$^1$, Marcin Skowron$^2$, Georgios Paltoglou$^3$, and Janusz A. Ho{\l}yst$^1$ } \affiliation{$^1$Faculty of Physics, Center of Excellence for Complex Systems Research, Warsaw University of Technology, Koszykowa 75, PL-00-662 Warsaw, Poland\\
$^2$ Interaction Technologies Group, Austrian Research Institute for Artificial Intelligence, Freyung 6/3/1a, A-1010 Vienna, Austria\\ 
$^3$ School of Computing and Information Technology, University of Wolverhampton, Wulfruna Street, Wolverhampton WV1 1SB, UK. 
}
\date{\today}
\begin{abstract}
We analyze emotionally annotated massive data from IRC (Internet Relay Chat) and model the dialogues between its participants by assuming that the driving force for the discussion is the entropy growth of emotional probability distribution. This process is claimed to be correlated to the emergence of the power-law distribution of the discussion lengths observed in the dialogues. We perform numerical simulations based on the noticed phenomenon obtaining a good agreement with the real data. Finally, we propose a method to artificially prolong the duration of the discussion that relies on the entropy of emotional probability distribution.     
\end{abstract} 
 \maketitle

\section{Introduction}
The extensive records of data opened new possibilities of examining communication between humans ranging from face-to-face encounters \cite{ftf1,ftf2,ftf3,ftf4}, through mobile telephone calls \cite{cphones1,cphones2}, surface-mail \cite{mail} short messages \cite{sms} to typical Internet activities such as e-mail correspondence \cite{email}, bulletin board system (BBS) dialogues \cite{bbs}, forum postings \cite{forum} or Twitter microblogging \cite{tweet}.  

Communication and its evolution is one of the key aspects of a modern life, which in an overwhelming part is governed by the circulation of information. In the most fundamental part, the communication is based on a {\it dialogue} - an exchange of information and ideas between two people \cite{dial}. Assuming an ideal situation, if the highest priority would be given to {\it acquiring} certain information, from a layman point of view the dialogue should be free from any additional components that could restrain conversation's participants from achieving the common goal. In reality, it is extremely difficult to model the dialogue complexity which, among others, would need to consider the dialogues' semantic, pragmatic, social and emotional context sequences of turn-taking \cite{turn1,turn2}, let alone its attentive \cite{lan_att} or contextual \cite{lan_context} layers.

As compared to the off-line communication, the exchange of information in the Internet is claimed to be more biased toward the emotional aspect \cite{hate}. It can be explained by a online disinhibition effect \cite{anon} --- the sense of anonymity that almost all Internet users possess while submitting their opinions on various fora or blogs. Nevertheless, it is the very Internet that gives the opportunity to acquire massive data, thus making it possible to perform a credible statistical analysis of common habits in communication. As the recent research shows, it is already possible to spot certain phenomena of the Internet discussion participants while looking just at the emotional content of their posts \cite{bosa1,frank,bosa2,ania1,ania2}. One of them is the collective emotional behavior \cite{ania1}, the other is clear correlation between the length of discussion and its emotional content \cite{ania1,ania2}.  

In this paper we argue that a simple physical approach based on the observation of entropy of emotional probability distribution during the conversation can serve as an indicator of a discussion about to finish. This process is claimed to be correlated to the emergence of the power-law distribution of the discussion length and serves as a key idea for the numerical simulations of the dialogues. The paper is organized as follows: Section \ref{sec:data} gives a brief description of the used data as well as of the emotional classification method, Section \ref{sec:comm_feat} presents our observations regarding the discussion length distribution, equalization of the emotional probabilities and entropy growth, in Section \ref{sec:sim_des} we show the description of simulations rules which results are given in Section \ref{sec:simres}. Finally, Section \ref{sec:app} describes a potential application of the observed phenomenon. 

\begin{figure}[!hb]
\centering
\includegraphics[width=0.8\columnwidth]{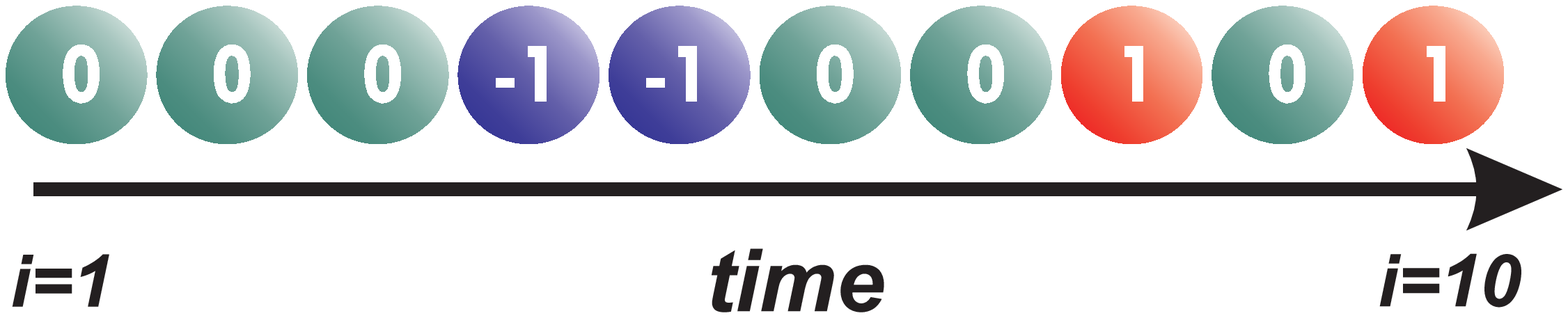}
\caption{(Color online) An exemplary dialogue of $L=10$ comments.}
\label{fig:chain}
\end{figure}

\begin{figure*}[!ht]
\centering
\includegraphics[width=\textwidth]{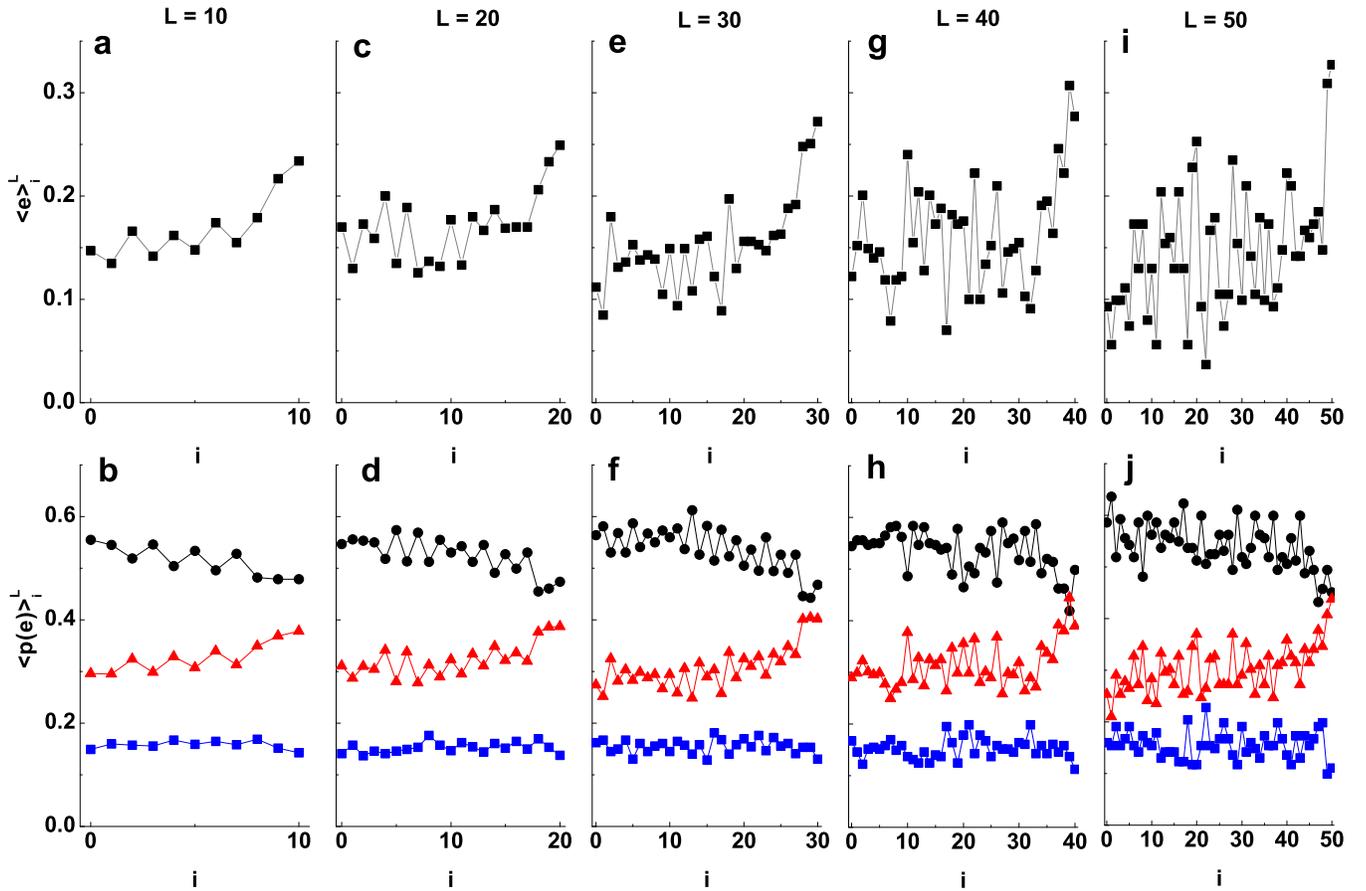}
\caption{(Color online) Average emotional value $\langle e \rangle^L_i$ (panels a, c, e, g, i) and average emotional probabilities $\langle p(-) \rangle^L_i$ (squares), $\langle p(0) \rangle^L_i$ (circles), $\langle p(+) \rangle^L_i$ (triangles) in the $i$-th timestep (panels b, d, f, h, j) for dialogues of specific length $L=10$ (a and b), $L=20$ (c and d), $L=30$ (e and f), $L=40$ (g and h) and $L=50$ (i and j).}
\label{fig:IRC_evol}
\end{figure*}

\begin{figure}[!b]
\centering
\includegraphics[width=\columnwidth]{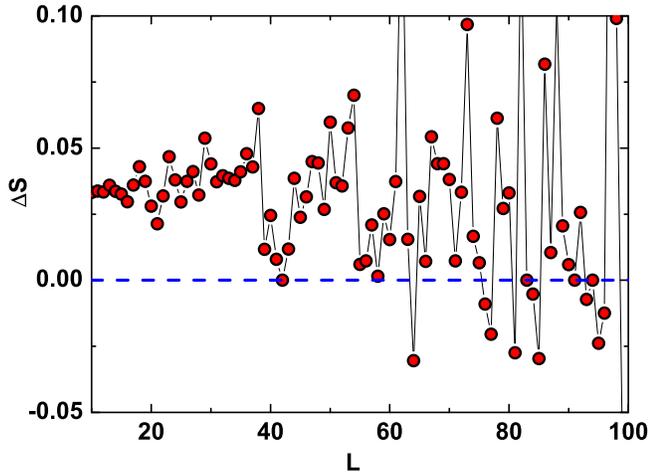}
\caption{(Color online) Difference between terminal and initial entropy value $\Delta S$ versus the dialogue length $L$.}
\label{fig:IRC_ent}
\end{figure}

\section{Data description}\label{sec:data}
As a source of data for analysing online dialogues we chose the Internet Relay Chat (IRC) \cite{irc} logs. Some of the the major IRC channels are being automatically archived by the channel operators, the logs are often accessible to a general public, and include the records of real-time, chat-like communication between numerous participants. The presented analysis is limited only to one of the channels, namely \texttt{\#ubuntu} \cite{ubuntu} in the period  1st January 2007 - 31st December 2009. In this work we focused on  dialogues that included only two participants. The final output, after several levels of data processing (for details see Appendix \ref{sec:dialext}) consists of $N=93329$ dialogues with the length $L$ between $L_{min}=11$ and $L_{max}=339$ each. Each dialogue can be represented as a chain of messages  (see Fig. \ref{fig:chain}) where all odd posts are submitted by one user and all even by another one.

\begin{figure*}[!t]
\centering
\includegraphics[width=\textwidth]{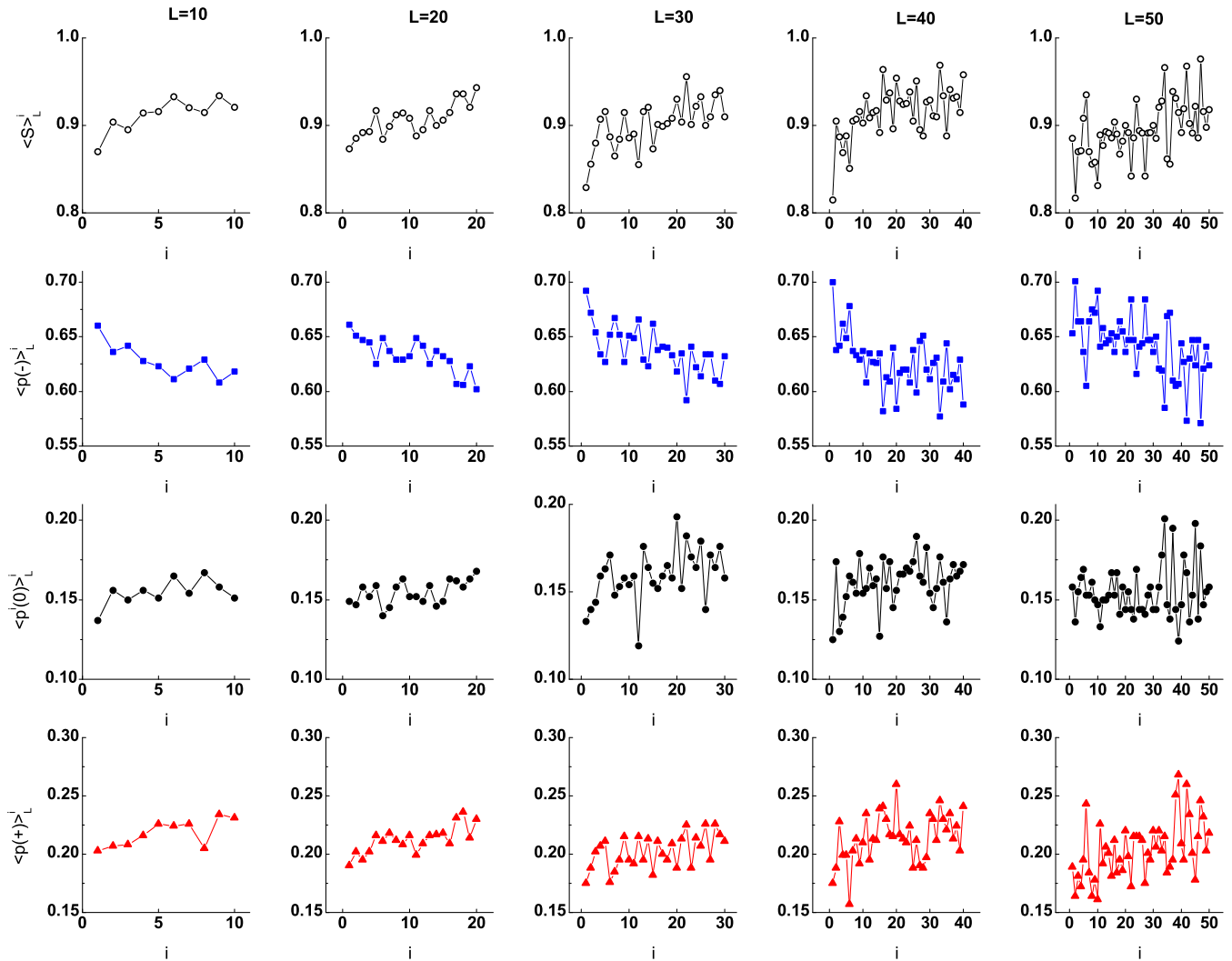}
\caption{(Color online) Entropy $S^{sh}_i$ of the average emotional probabilities distribution $\langle p(e) \rangle^L_i$ (topmost row) and average emotional probabilities $\langle p(-) \rangle^L_i$ (squares) $\langle p(0) \rangle^L_i$ (circles) and $\langle p(+) \rangle^L_i$ (triangles) in the $i$-th timestep for BBC Forum discussions of specific length $L=10$ (first column), $L=20$ (second column), $L=30$ (third column), $L=40$ (fourth column) and $L=50$ (fifth column).}
\label{fig:BBC_ent}
\end{figure*}

The emotional classifier program that was used to analyze the emotional content of the discussions is based on a machine-learning (ML) approach. The algorithm functions in two phases: during the training phase, it is provided with a set of documents classified by humans for emotional content (positive, negative or objective) from which it learns the characteristics of each category. Then, during the application phase, the algorithm applies the acquired sentiment classification knowledge to new, unseen documents. In our analysis, we trained a hierarchical Language Model \cite{hLM,lm2} on the Blogs06 collection \cite{blogs} and applied the trained model to the extracted IRC dialogues, during the application phase. The algorithm is based on a two-tier solution, according to which a post is initially classified as objective or subjective and in the latter case, it is further classified in terms of its polarity, i.e., positive or negative. Each level of classification applies a binary Language Model \cite{lm1,lm2}. Posts are therefore annotated with a single value $e =-1, 0$ or $1$ to quantify their emotional content (to be more precise - their valence \cite{valence}) as negative, neutral or positive, respectively.

\section{Common features}\label{sec:comm_feat}
The obtained dialogues have been divided into groups of constant dialogue length $L$. For such data we follow the evolution of mean emotional value $\langle e \rangle^L_i$ and average emotional probabilities $\langle p(e) \rangle^L_i$ ($\langle e \rangle^L_i$. In both cases the $\langle ... \rangle^L_i$ symbol indicates taking all dialogues with a specific length $L$ and averaging over all comments with number $i$, thus, for example, $\langle p(-) \rangle^L_i$ is the probability that at the position $i$ in all dialogues of length $L$ there is a negative statement. The characteristic feature observed regardless of the dialogue length is that the $\langle e \rangle^L_i$ at the end of the dialogue is higher than at the beginning (upper row in Fig. \ref{fig:IRC_evol}). In fact, there is especially a rapid growth close the very end of the dialogue.

\begin{figure}[!ht]
\centering
\includegraphics[width=\columnwidth]{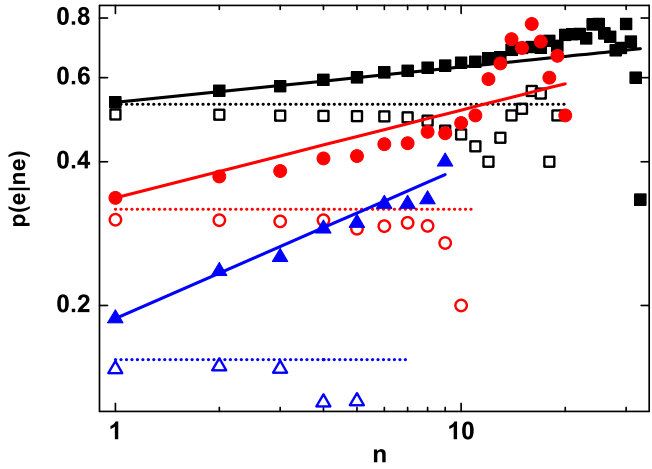}
\caption{(Color online) Conditional probability $p(e|ne)$ of consecutive emotional post of the same sign versus the size $n$.  Full triangles, squares and circles are data points (respectively: negative, neutral and positive messages), empty symbols are shuffled data, solid lines come from Eq. (\ref{eq:alfa}) and dotted lines represent relation $p(e|ne)=p(e)$.}
\label{fig:IRC_clust}
\end{figure}

The direct reason for such behavior is shown in the bottom row of Fig. \ref{fig:IRC_evol}, which presents the evolution of the average emotional probabilities $\langle p(-) \rangle^L_i$, $\langle p(0) \rangle^L_i$ and $\langle p(+) \rangle^L_i$. The observations can be summarized in the following way:
\begin{itemize}
\item the negative emotional probability $\langle p(-) \rangle^L_i$ remains almost constant,
\item $\langle p(+) \rangle^L_i$ increases and $\langle p(0) \rangle^L_i$ has an opposite tendency,
\item $\langle p(+) \rangle^L_i$ and $\langle p(0) \rangle^L_i$ tend to equalize in the vicinity of dialogue end.
\end{itemize}

\begin{figure*}[!ht]
\centering
\includegraphics[width=0.75\textwidth]{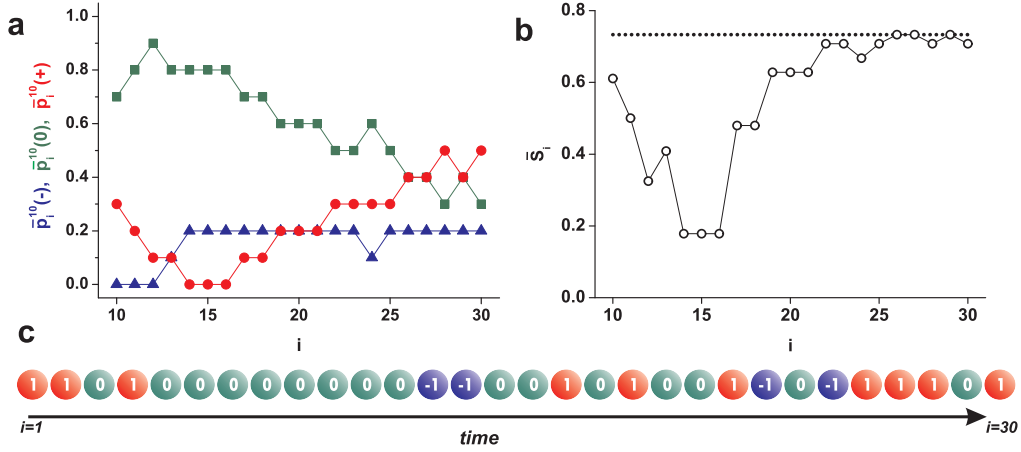}
\caption{(Color online) (a) Probabilities of specific valence $\bar{p}_i^M(-)$ (triangles), $\bar{p}_i^M(0)$ (squares) and $\bar{p}_i^M(+)$ (circles) in the $i$-th time window given by Eq. (\ref{eq:pmov}) for the exemplary dialogue shown in panel (c). (b) Entropy $\bar{S}_i$ in the $i$-th time window defined by Eq. (\ref{eq:Sp}) for the exemplary dialog shown in panel (c). The dotted line marks the maximal value of entropy in Eq. (\ref{eq:Sp}) i.e., $\bar{S}^{max}_i=0.4 \ln 2.5 \approx 0.73$.}
\label{fig:dial}
\end{figure*}

Other manifestation of the system's features can be spotted by examining the level of the entropy $S$ of the emotional probabilities $\langle p(e) \rangle^L_i$. Entropy or other information theoretic quantities as mutual information \cite{mi}, Kullback-Leiber divergence \cite{div} or Jensen-Shannon divergence \cite{div} have been already used to quantify certain aspects of human mobility \cite{mobility}, semantic resemblance or flow between Wikipedia pages \cite{masucci_pre,masucci_plos} or correlations between consecutive emotional posts \cite{acta}. Moreover, basing on entropy, it has also been shown how the coherent structures in the e-mail dialogues arise \cite{email} or how to predict conversation patterns in face-to-face meetings \cite{ftf3}. In this paper, the entropy is used after Shannon's definition \cite{shannon}, i.e.,
\begin{equation}
S^{sh}_i = -\sum_{e={-1,0,1}}\langle p(e) \rangle^L_i \ln \langle p(e) \rangle^L_i.
\end{equation}
Here, taking into account the fact that $\langle p(-) \rangle^L_i$ is constant in the course of dialogue, we paid attention only to $\langle p(+) \rangle^L_i$ and $\langle p(0) \rangle^L_i$, thus the observed entropy had a form of
\begin{equation}
S_i = -\left[ \langle p(0) \rangle^L_i \ln \langle p(0) \rangle^L_i + \langle p(+) \rangle^L_i\ln \langle p(+) \rangle^L_i \right].
\label{eq:S}
\end{equation}

Plotting the difference between terminal and initial entropy $\Delta S$ versus the length of the dialogue $L$ it is possible to see that for the dialogues up to $L \approx 50$ this difference is always above zero (see Fig. \ref{fig:IRC_ent}). It implies a following likely scenario for the dialogue: it evolves in the direction of growing entropy. In the beginning of the dialogue, the probabilities $\langle p(0) \rangle^L_i$ and $\langle p(+) \rangle^L_i$ are separated from each other, contributing to low value of initial entropy $S_p$. However, then the entropy grows, the probabilities $\langle p(0) \rangle^L_i$ and $\langle p(+) \rangle^L_i$ equalize leading to high value entropy (i.e., higher than the initial one) at the end of the dialogue.   

However, it is essential to notice that the observed behavior in the IRC data is only one of the possible scenarios of the more general phenomenon of the principle of maximum entropy \cite{jaynes}, governing also certain aspects of biological \cite{williams} or social systems \cite{johnson} (at the level of social networks). To be more precise, we performed an analysis analogous to this for the IRC data with respect to emotionally annotated dataset from the BBC Forum (see \cite{ania1} and \cite{ania2}) consisting of over $2\times10^6$ comments and almost $10^5$ discussions. In this case each discussion was treated as a natural "dialogue", although it usually consisted of more than 2 users communicating to each other. Following the line of thought presented for IRC data we grouped all discussion of constant length and calculated the quantities $\langle p(-) \rangle^L_i$, $\langle p(0) \rangle^L_i$, $\langle p(+) \rangle^L_i$ and $S^{sh}_i$. The results, shown in Fig. \ref{fig:BBC_ent}, bear close resemblance to those obtained for IRC data: one can clearly see that while the negative component decreases, the positive and objective (partially) ones increase. It has an instant effect on the value of entropy which grows during the evolution of the discussion (topmost row in Fig. \ref{fig:BBC_ent}). The main difference between IRC and BBC Forum results concerns the component whose value decreases during the discussion evolution: for IRC it is the $\langle p(0) \rangle^L_i$ while for BBC Forum - $\langle p(-) \rangle^L_i$. It is directly connected to the fact that the above mentioned components play the role of "discussion fuel" \cite{ania1} propelling thread's evolution. BBC Forum data come from such categories as "World News" and "UK News" and as such may lead the discussion participants to place comments of very negative valence. On the other hand \texttt{\#ubuntu} IRC channel servers rather as a source of professional help which is normally expressed in terms of neutral dialogue. As the discussion lasts, the topic dilutes (BBC Forum) or the problem is being solved (IRC) and the dominating component dies out leading to maximization of entropy.

There is also another process taking place in the system in question that displays a non-trivial behavior. As shown previously in \cite{ania1}, we can talk about grouping of similarly emotional messages. To quantify the persistence of a specific emotion one can consider the conditional probability $p(e|ne)$ that after $n$ comments with the same emotional valence the next comment has the same sign. As it easy to prove, if $e$ were an identical and independently distributed (i.i.d.) variable the conditional probability $p(e|ne)$ should be independent of $n$ and equal to $p(e)$, i.e., the probability of a specific emotion in the whole dataset (see Table \ref{tab:IRC_prop}) In the case of the IRC data, the analysis shows (see Fig. \ref{fig:IRC_clust}) that $p(e|ne)$ is well approximated by 
\begin{equation}
p(e|ne) = p(e|e)n^{\alpha}.
\label{eq:alfa}
\end{equation}
where $p(e|e)$ is the conditional probability that two consecutive messages have the same emotion. The discrepancy between the data and the relation obtained by random insertion of emotional comments (see open symbols in Fig. \ref{fig:IRC_clust}) is significant. The exponents $\alpha$ and the conditional probabilities $p(e|e)$ are gathered in Table \ref{tab:IRC_prop}.

\begin{table}[htb]
\setlength{\tabcolsep}{8.5pt}
\centering
\begin{tabular}{lrrr}
\hline
\hline
emotion sign & $p(e)$ & $p(e|e)$ & $\alpha_e$\\
\hline
positive ($e=1$) & 0.318 & 0.34 & 0.18\\
neutral ($e=0$) & 0.528 & 0.53 & 0.07\\
negative ($e=-1$) & 0.154 & 0.19 & 0.31\\
\hline
\hline
\end{tabular}
\caption{Fundamental properties of dialogue data: probabilities of specific emotion $p(e)$, conditional probabilities $p(e|e)$ and scaling exponents for the power-law cluster growth $\alpha_e$.}
\label{tab:IRC_prop}
\end{table}

\begin{figure*}[!ht]
\centering
\includegraphics[width=0.85\textwidth]{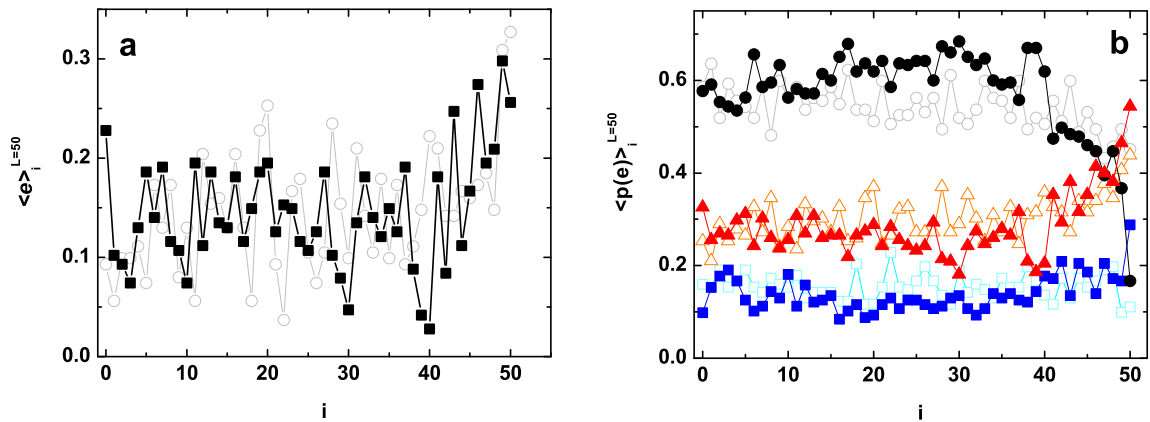}
\caption{(Color online) Comparison of average emotional value $\langle e \rangle$ (panel a) and probability of specific emotion (panel b, $\langle p(-) \rangle^{L=50}_i$ - squares, $\langle p(0) \rangle^{L=50}_i$ - circles, $\langle p(+) \rangle^{L=50}_i$ - triangles) for simulations performed according to the procedure presented in Sec. \ref{sec:sim_des} (full symbols) and for real data (empty symbols) for dialogue length $L=50$. The real data shown are identical with those shown in Fig. \ref{fig:IRC_evol}i and Fig. \ref{fig:IRC_evol}j.}
\label{fig:IRC_comp}
\end{figure*}

\section{Simulation description}\label{sec:sim_des}

The methodology described above proves to be successful in finding the prominent characteristic of the data in question, however it is rather useless if one would like to perform the simulations of the dialogues. It is crucial to choose other way for calculating the average emotional probabilities "on the fly" and, using the results, decide on the further dialogue evolution. Thus, we decided to work with moving time window, i.e, the probabilities of the specific valences in the $i$-th timestep are
\begin{equation}
\left\{
\begin{array}{l}
\bar{p}_i^M(+) = \frac{1}{M}\sum_{j=1}^{j=M} \delta_{e(i-j),+1},\\
\bar{p}_i^M(0) = \frac{1}{M}\sum_{j=1}^{j=M} \delta_{e(i-j),0},\\
\bar{p}_i^M(-) = \frac{1}{M}\sum_{j=1}^{j=M} \delta_{e(i-j),-1},
\end{array}
\right.
\label{eq:pmov}
\end{equation}
for $i \ge M$, where $\delta$ is the Kronecker delta symbol and $M$ is the size of the window. Consequently, entropy $S_i$ is also calculated using the probabilities $\bar{p}^M_i (+)$ and $\bar{p}^M_i (0)$ as 
\begin{equation}
\bar{S}_i = -\left[ \bar{p}^M_i(0)\ln \bar{p}^M_i(0) + \bar{p}^M_i(+)\ln \bar{p}^M_i(+) \right].
\label{eq:Sp}
\end{equation}
expressing in fact the entropy in the $i$-th time window. The practical way of application is shown in Fig. \ref{fig:dial} for a dialogue of $L=30$ comments. In this case the size of the time window is set to $M=10$.

\begin{figure*}[!ht]
\centering
\includegraphics[width=0.85\textwidth]{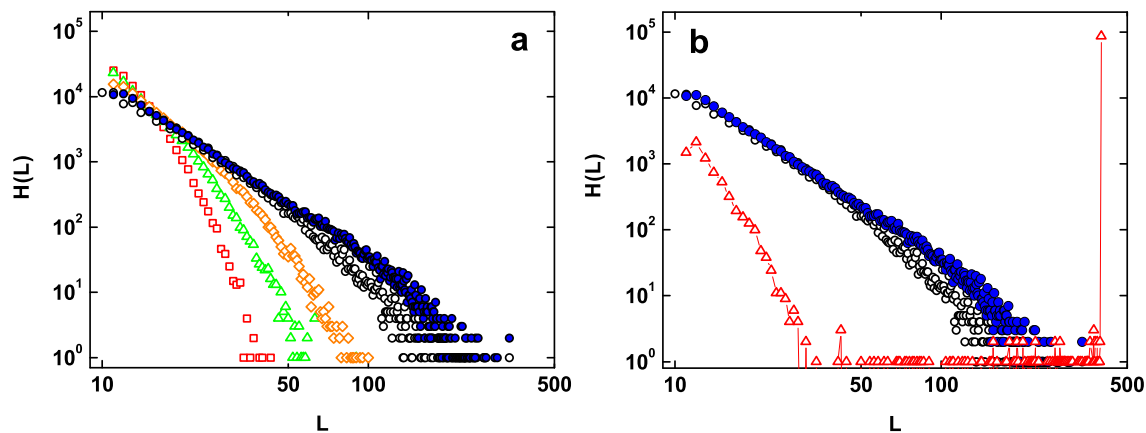}
\caption{(Color online) (a) Dialogue length distribution $H(L)$ for real data (empty circles) and simulations for different values of the initial entropy threshold $S_T$ parameter: $S_T=0.1$ (empty squares), $S_T=0.5$ (empty triangles), $S_T=0.6$ (empty diamonds) and $S_T=0.67$ (filled circles). (b) Dialogue length distribution $H(L)$ for: real data (empty circles), simulations with $S_p=0.67$ (filled circles) and simulations with $S_T=0.67$ and insertion of the additional neutral comments (empty triangles). The solid line is for visual guidance.}
\label{fig:IRC_length}
\end{figure*}

The data-driven facts presented in the previous section lie at the basis of the simulation of dialogues in IRC channels data. The key point treated as an input parameter for this model is the observation of the preferential attraction of consecutive emotional messages. This idea "runs" the dialogue, whereas the discussion is terminated once the difference between the entropy in the given moment and its initial value exceeds certain threshold. Those features are implemented in the following algorithm: 
\begin{enumerate}
\item[(1)] start the dialogue by drawing the first emotional comment with probability $p(e)$, 
\item[(2)] set the next comment to have emotional valence $e$ of the previous comment with probability $p(e|ne)=p(e|e)n^{\alpha_e}$
\item[(3)] if the drawn probability is higher than $p(e|ne)$, set the next comment one of two other emotional values (i.e., if the original $e=1$, then the next comment valence is 0 with probability $p(0)/[p(0)+p(-)]$ or -1 with probability $p(-)/[p(0)+p(-)]$)
\item[(4)] if the difference between entropy in this time-step and the initial entropy is higher than threshold level $\Delta S=0.05$ terminate the simulation, otherwise go to point (2). 
\end{enumerate}
The observed valence probabilities in this simulation are always calculated using quantities in a moving time window given by Eqs. (\ref{eq:pmov}-\ref{eq:Sp}) with $M=10$.

There is another crucial parameter connected to the simulation process, i.e., the initial entropy threshold $S_T$. When time-step $i=M$ is reached, the entropy $\bar{S}_i$ is calculated for the first time and then decision is taken: if $\bar{S}_M < S_T$ the simulation runs further, otherwise it is cancelled and repeated. The total number of successfully simulated dialogues is equal to this observed in the real data.

\section{Simulation results}\label{sec:simres}
Figure \ref{fig:IRC_comp} shows a comparison of the average emotional value $\langle e \rangle^L_i$ and average emotional probabilities $\langle p(e) \rangle^L_i$ for the real data and simulations performed according to the algorithm described in the previous section for dialogues of length $L=50$. As one can see the plots bear close resemblance apart from only one detail, i.e., the rising value for the $\langle p(-) \rangle^L_i$ close to the end of the dialogue.

Moreover, the simulation strongly depends on the exact value of the initial entropy threshold $S_T$ which can be clearly seen in Fig. \ref{fig:IRC_length}a, where the dialogue length distribution is presented. If the $S_T$ is restricted to values between 0.1-0.5 (empty squares and triangles) the distribution of dialogue lengths is exponential and does not follow the one observed in the real data (empty circles). Higher values of $S_T$ ($S_T=0.6$, empty diamonds) shift the curve closer to the data points, nevertheless the character is still exponential. It is only after tuning the $S_T$ parameter to 0.67 that the results obtained from the simulations (full circles) are qualitatively comparable with the real data.

\section{Application}\label{sec:app}
It is possible to consider a direct application of the above described model for changing the "trajectory" of the dialogue. For example let us assume that a dialogue system \cite{ds1,ds2,ds3} is included as part of the conversation and that its task is to prolong the discussion. In such situation, the system that could rely on the above presented properties  would attempt to detect any signs indicating that the dialogue might come to an end and react against it. According to observations presented in section \ref{sec:comm_feat} a marker for such event should be the growth of the entropy. In other words the dialogue system should prevent an increase of the entropy in the consecutive time-steps. 

In the described case, such action would be an equivalent to an insertion of an objective comment. In this way, an equalization between $\bar{p}^M_i(+)$ and $\bar{p}^M_i(0)$ is prevented and dialogue can last further. An implementation of this rule is presented in Fig. \ref{fig:IRC_length}b, where one can compare the real data (again empty circles), a simulation including the entropy-growth rule (again full circles) and a simulation following the insertion of objective comments (empty triangles). While there is a drop-down in the numbers for the small dialogue lengths, the vast majority of the dialogues has the maximal length (a point in the top-right corner). In this way the insertion of the objective comments is in line with the expected idea of dialogue prolonging. 
\section{Conclusions}
Analysis performed on the emotionally annotated dialogues extracted from IRC data demonstrate that following such simple metrics as probability of specific emotion can be useful to predict the future evolution of the discussion. Moreover, all the analysed dialogues share the same property, i.e., the tendency to evolve in the direction of a growing entropy. Those features, combined together with the observations regarding the preferential growth of clusters, are sufficient to reproduce the real data by a rather straightforward simulation model.  
In the paper, we also proposed a procedure to directly apply the observed rules in order to modify the way the dialogue evolves. It appears, that insertion of comments with emotion that initially had dominated and then started to vanish prolongs the discussion by lowering the entropy value. Those observations may be of help for designing the next generation of interactive software tools\cite{app1,app2,app3} intended to support e-communities by measuring various features of their interactions patterns, including their emotional state at the individual, group and collective levels.

\begin{acknowledgments}
This work was supported by a European Union grant by the 7th Framework Programme, Theme 3: Science of complex systems for socially intelligent ICT. It is part of the CyberEmotions (Collective Emotions in Cyberspace) project (contract 231323). J.S. and J.A.H. also acknowledge support from the European COST Action MP0801 Physics of Competition and Conflicts as well as from the Polish Ministry of Science Grants Nos. 1029/7.PR UE/2009/7 and 578/N-COST/2009/0.
\end{acknowledgments}

\appendix
\section{Dialogue extraction method}\label{sec:dialext}
\begin{table*}[htb]
\setlength{\tabcolsep}{8.5pt}
\centering
\begin{tabular}{llllc}
\hline
\hline
Original data & User-to-user info & Output 1 & Output 2 & Final output\\
\hline
1 $[00:03]$ $\langle 20422 \rangle$ 1 	& $[00:03]$ $\langle20442\rangle$ 				& 	  						&  							& Dialogue 1\\
2 $[00:04]$ $\langle55\rangle$ 1 	& $[00:04]$ $\langle55\rangle \rightarrow \langle 20442 \rangle$& $\langle55\rangle \rightarrow \langle20442\rangle$ 1 	& $\langle55\rangle \rightarrow \langle20422\rangle$ 1 & $\langle55\rangle \leftrightarrow \langle20422\rangle$\\
3 $[00:05]$ $\langle20422\rangle$ 0 	& $[00:05]$ $\langle20442\rangle \rightarrow \langle55\rangle$ 	& $\langle20442\rangle \rightarrow \langle55\rangle$ 0	& $\langle20422\rangle \rightarrow \langle55\rangle$ 0 & 1\\
4 $[00:05]$ $\langle55\rangle$ -1 	& $[00:05]$ $\langle55\rangle \rightarrow \langle20442\rangle$ 	& $\langle55\rangle \rightarrow \langle20442\rangle$ -1 & $\langle55\rangle \rightarrow \langle20422\rangle$ -1 & 0\\
5 $[00:08]$ $\langle20422\rangle$ 1 	& $[00:08]$ $\langle20422\rangle \rightarrow \langle55\rangle$ 	& $\langle20442\rangle \rightarrow \langle55\rangle$ 1 	& $\langle20422\rangle \rightarrow \langle55\rangle$ 1 & -1\\
6 $[00:08]$ $\langle55\rangle$ 0 	& $[00:08]$ $\langle55\rangle \rightarrow \langle20442\rangle$ 	& $\langle55\rangle \rightarrow \langle20442\rangle$ 0 	& $\langle55\rangle \rightarrow \langle20442\rangle$ 0 & 1\\
7 $[00:09]$ $\langle27\rangle$ 0 	& $[00:09]$ $\langle27\rangle \rightarrow \langle20442\rangle$ 	& $\langle27\rangle \rightarrow \langle20442\rangle$ 0 	& $\langle27\rangle \rightarrow \langle20442\rangle$ 0 & 0\\
8 $[00:13]$ $\langle20422\rangle$ 0	& $[00:13]$ $\langle 20422\rangle$ 				& $\langle20442\rangle \rightarrow \langle27\rangle$ 0 	& $\langle20422\rangle \rightarrow \langle27\rangle$ 0	& Dialogue 2\\
9 $[00:13]$ $\langle2\rangle$ -1	& $[00:13]$ $\langle 2\rangle$ 					&  							&  							& $\langle20422\rangle \leftrightarrow \langle27\rangle$\\
10 $[00:14]$ $\langle20422\rangle$ -1 	& $[00:14]$ $\langle20422\rangle \rightarrow \langle20442\rangle$ & $\langle20442\rangle \rightarrow \langle27\rangle$ -1 &  							& 0\\
11 $[00:14]$ $\langle20422\rangle$ 0 	& $[00:14]$ $\langle 20422\rangle$ 				& $\langle20442\rangle \rightarrow \langle27\rangle$ 0	&  							& 0\\
12 $[00:59]$ $\langle171\rangle$ -1 	& $[00:59]$ $\langle171\rangle \rightarrow \langle13692\rangle$ & $\langle171\rangle \rightarrow \langle13692\rangle$ -1& $\langle171\rangle \rightarrow \langle13692\rangle$ 0 & Dialogue 3\\
13 $[00:59]$ $\langle171\rangle$ 1 	& $[00:59]$ $\langle171\rangle \rightarrow \langle13692\rangle$ & $\langle171\rangle \rightarrow \langle13692\rangle$ 1	&  							& $\langle171\rangle \leftrightarrow \langle13692\rangle$\\
14 $[00:59]$ $\langle171\rangle$ 0 	& $[00:59]$ $\langle171\rangle \rightarrow \langle13692\rangle$ & $\langle171\rangle \rightarrow \langle13692\rangle$ 0	&  							& 0\\
15 $[01:00]$ $\langle171\rangle$ 1 	& $[01:00]$ $\langle171\rangle \rightarrow \langle13692\rangle$ & $\langle171\rangle \rightarrow \langle13692\rangle$ 1	&  							& 0\\
16 $[01:00]$ $\langle13692\rangle$ 0 	& $[01:00]$ $\langle13692\rangle$ 				& $\langle13692\rangle \rightarrow \langle171\rangle$ 0	& $\langle13692\rangle \rightarrow \langle171\rangle$ 0 & 1\\
17 $[01:01]$ $\langle171\rangle$ 1 	& $[01:01]$ $\langle171\rangle \rightarrow \langle13692\rangle$ & $\langle171\rangle \rightarrow \langle13692\rangle$ 1	& $\langle171\rangle \rightarrow \langle13692\rangle$ 1 & 1\\
18 $[01:01]$ $\langle171\rangle$  1 	& $[01:01]$ $\langle171\rangle \rightarrow \langle13692\rangle$ & $\langle171\rangle \rightarrow \langle13692\rangle$ 1	&  							& 1\\
19 $[01:01]$ $\langle13692\rangle$ 1 	& $[01:01]$ $\langle13692\rangle$ 				& $\langle13692\rangle \rightarrow \langle171\rangle$ 1	& $\langle13692\rangle \rightarrow \langle171\rangle$ 1	& 1\\
20 $[01:01]$ $\langle171\rangle$ 1 	& $[01:01]$ $\langle171\rangle$ 				& $\langle171\rangle \rightarrow \langle13692\rangle$ 1	& $\langle171\rangle \rightarrow \langle13692\rangle$ 1	& -1\\
21 $[01:02]$ $\langle171\rangle$ 1 	& $[01:02]$ $\langle171\rangle \rightarrow \langle13692\rangle$ & $\langle171\rangle \rightarrow \langle13692\rangle$ 1	& 							& 1\\
22 $[01:02]$ $\langle171\rangle$ 1 	& $[01:02]$ $\langle171\rangle \rightarrow \langle13692\rangle$ & $\langle171\rangle \rightarrow \langle13692\rangle$ 1	& 							& -1\\
23 $[01:02]$ $\langle13692\rangle$ 1 	& $[01:02]$ $\langle13692\rangle$ 				& $\langle13692\rangle \rightarrow \langle171\rangle$ 1	& $\langle13692\rangle \rightarrow \langle171\rangle$ 1	& 1\\
24 $[01:02]$ $\langle13692\rangle$ 0 	& $[01:02]$ $\langle13692\rangle$ 				& $\langle13692\rangle \rightarrow \langle171\rangle$ 0	& 							& 		\\
25 $[01:02]$ $\langle171\rangle$ -1 	& $[01:02]$ $\langle171\rangle \rightarrow \langle13692\rangle$ & $\langle171\rangle \rightarrow \langle13692\rangle$ -1& $\langle171\rangle \rightarrow \langle13692\rangle$ -1& 		\\
26 $[01:03]$ $\langle13692\rangle$ 1 	& $[01:03]$ $\langle13692\rangle$ 				& $\langle13692\rangle \rightarrow \langle171\rangle$ 1	& $\langle13692\rangle \rightarrow \langle171\rangle$ 1	& 		\\
27 $[01:03]$ $\langle13692\rangle$ -1 	& $[01:03]$ $\langle13692\rangle$ 				& $\langle13692\rangle \rightarrow \langle171\rangle$ -1& 							& 		\\
28 $[01:03]$ $\langle13692\rangle$ 1 	& $[01:03]$ $\langle13692\rangle$ 				& $\langle13692\rangle \rightarrow \langle171\rangle$ 1	& 							& 		\\
29 $[01:03]$ $\langle171\rangle$ -1 	& $[01:03]$ $\langle171\rangle$ 				& $\langle171\rangle \rightarrow \langle13692\rangle$ -1& $\langle171\rangle \rightarrow \langle13692\rangle$ -1& \\
20 $[01:03]$ $\langle13692\rangle$ 1 	& $[01:03]$ $\langle13692\rangle$ 				& $\langle13692\rangle \rightarrow \langle171\rangle$ 1 & $\langle13692\rangle \rightarrow \langle171\rangle$ 1	& \\
\hline
\hline
\end{tabular}
\caption{The process of dialogue extraction in the IRC channel data. Columns from the left show consecutive steps of the algorithm: first and second show the raw data, third is data after application of the searching procedure, fourth is data after averaging multiple posts from the same user and fifth column gives the final output. $[hh:mm]$ defines the timestamp in hours ($hh$) and minutes ($mm$), $\langle user\_id \rangle$ gives the id of the user that addresses the post, $\langle adressing\_user\_id \rangle \rightarrow \langle addresed\_user\_id \rangle$ gives the ids of both addressing and addressed users and value $\{-1,0,1\}$ shows the valence of the post.}
\label{tab:IRC_data}
\end{table*}

In total, we used 994 daily files with 4600 to 18000 utterances that share a format presented in the first column from the left in Table \ref{tab:IRC_data}: $post\_number$ $[timestamp]$ $\langle user\_id \rangle$ $sentiment\_class$ with the $sentimentclass$ $e=\{-1;0,1\}$ used as marker for the emotional valence through this study. Moreover, we could also use information that specifies which user communicates, i.e., directly addresses, another user (see second column in Table \ref{tab:IRC_data}, shown as $\langle addressing\_user\_id \rangle \rightarrow \langle addressed\_user\_id\rangle$). The discovery of the direct communication links between two users in the IRC channel was based on the discovery of another userID at the beginning of an utterance, followed by a comma or semicolon signs; a scheme commonly used in various multiple users communication channels.
However, one has to bear in mind that this kind of information can be sometimes incomplete, i.e., in many cases users do not explicitly specify the receiver of his/her post. Another issue that arises is that the data consist of several overlapping dialogues held simultaneously on one channel. It is also sometimes difficult to indicate the receiver of the message as only part of them are annotated with a user id they are dedicated to. We created an algorithm that addresses this issue. It consists of two different approaches:
\begin{enumerate}
\item[(a)] if user A addresses user B in some moment in time and later A writes consecutive messages without addressing anybody specific we assume that he/she is still having a conversation with B
\item[(b)] if user A addresses user B and then B writes a message without addressing anybody specific we assume that he/she is answering to A.
\end{enumerate}
The main parameter of such algorithm is the time $t$ in which the searching is being done; in our study we use $t=5$ minutes as the threshold value . An exemplary output from the algorithm is shown in the third column in Table \ref{tab:IRC_data}. In this way we are able to extract a set of dialogues from each of the daily files.
After processing the file according to above described rules another issue emerges: it often happens that a user gives a set of consecutive messages directed to one receiver (e.g, the 8th, 10th and 11th line in the third column in Table \ref{tab:IRC_data}). To create a standardize version of the dialogue (A to B, B to A, A to B and so on), we decided to accumulate the consecutive emotional messages of the same user, calculate the average value $\bar{e}$ in such series and then  transform it back into a three-sate value according to the formula
\begin{equation}
\left\{
\begin{array}{lcl}
e^{i}=-1 & & \bar{e} \in [-1;-\frac{1}{3}]\\
e^{i}=0 & & \bar{e} \in (-\frac{1}{3};\frac{1}{3})\\
e^{i}=1 & & \bar{e} \in [\frac{1}{3};1]\\
\end{array}
\right.
\end{equation}
In effect we obtain the set shown in the fourth column in Table \ref{tab:IRC_data}. The final step of the data preparation is to divide it into separate dialogues as shown in the 5th column in Table \ref{tab:IRC_data}. In total, the algorithm produces $N=93329$ dialogues with the length between $L=11$ and $L=339$ (all the dialogues with $L \le 10$ were omitted).

\end{document}